\documentclass[prb]{revtex4}
\usepackage{graphicx}
\usepackage{dcolumn}
\usepackage{bm}

\begin{document}

\preprint{APS/123-QED}

\title{Tunneling Violates Special Relativity}
\author{G\"unter Nimtz}
 \altaffiliation
{G.Nimtz@uni-koeln.de}
\affiliation{
II. Physikalisches Institut, Universt\"at zu K\"oln \\
Z\"ulpicherstrasse 77, 50937 K\"oln
}%


\date{\today}

\begin{abstract}
Experiments with evanescent modes and tunneling particles have shown that
i) their signal velocity may be faster than light,
ii) they are described by virtual particles,
iii)  they are nonlocal and act at a distance,
iv) experimental tunneling data of phonons, photons, and electrons display a universal scattering time at the
tunneling barrier front, and v) the properties of evanescent, i.e. tunneling modes is not compatible with the special theory of relativity.
\end{abstract}

\maketitle

\section{Introduction }

How much time does a wave packet spend in a potential barrier?
As shown in Table 1 the time varied from milliseconds to attoseconds by 15 orders of magnitude measured with different particles and barriers~\cite{NimtzM}. The time resulted in faster than light velocities since the velocity given by barrier length divided by the time exceeds c.  This tunneling time was found to be a universal property for phonons, photons, and electrons being only dependent on the wave packet frequency~\cite{NimtzM,Esposito}. Actually, it was found that the measured tunneling time is raised at the potential barrier boundary, whereas zero time is spent inside the barrier.

\begin{table*}
\begin{center}
\begin{tabular}{|l|l|c|c|c|}

\hline
\hline Tunneling barriers & Reference  & $\tau$   &  $T=1/\nu$  \\
\hline
\hline {\it frustrated total reflection}   & Ref.~\cite{Haibel} & 117\,ps & 120\,ps \\
\hline
\cline{2-4} {\it at double prisms}   & Ref.~\cite{Bal}  &  30\,fs &  11.3\,fs \\
\hline
\cline{2-4} {\it  }  & Ref.\cite{Mug}  & 87\,ps & 100\,ps  \\
\cline{2-4}
\hline
\hline {\it photonic lattice}  & Ref.\cite{Steinberg}. & 2.13\,fs & 2.34\,fs \\
\hline
\cline{2-4}{\it } &  Ref.\cite{Krausz}  &  2.7\,fs  &  2.7\,fs  \\
\hline
\hline {\it undersized waveguide} & Ref.\cite{Enders} & 130\,ps & 115\,ps \\
\hline
\hline {\it electron field-emission tunneling} & Ref.~\cite{Sek}  & 7\,fs & 6\,fs \\
\hline
\hline {\it electron ionization tunneling} & Ref.~\cite{Keller}  & $\leq$ 6\,as &  ?\,as \\
\hline
\hline {\it acoustic (phonon) tunneling } & Ref.~\cite{Yang} & 0.8\,$\mu$s &  1\,$\mu$s  \\
\hline
\hline {\it acoustic (phonon) tunneling } & Ref.~\cite{Robertson} & 0.9\,ms &  1\,ms  \\
\hline
\hline
\end{tabular}
\end{center}
\caption{Experimental $\tau$ and T = $1/\nu$ transmission time (\emph{tunneling time}), of photons, phonons, and electrons. It was observed that the measured tunneling time $\tau$ equals approximately the wave packet's oscillation time T. Notice, the measured barrier traversal time $\tau$ differ about 15 orders of magnitude depending on the wave packet energy $\nu$h ~\cite{NimtzM}.}
\end{table*}

This brief article represents a summary of several extensive review articles published in the last ten years~\cite{NimtzM,NimtzH,Nimtz,Nimtza}. The impetus for this brief note is a recent review \emph{on slow and fast light}, which omitted reporting on the velocity of evanescent modes and tunneling particles~\cite{Boyd}.

Photonic tunneling, i.e. propagation of the so-called evanescent
modes in classical optics is studied much easier than particles
like electrons. Thus the first superluminal tunneling velocity was measured with
microwaves in 1992~\cite{Enders}. Later
experimental results of frustrated total internal reflection
and of periodic dielectric lattices in digital optical fiber communication confirmed the superluminal signal velocity of tunneling modes~\cite{Nimtz,Nimtza,Longhi,Longhi2}.

The signal velocity is the relevant velocity studied in this report. It is the velocity of the transmitted cause, i.e. of the information. A signal may be attenuated (from forte to piano) but not deformed. For example in a digital pulse, the temporal half width equals the number of bits and is attenuated  along a wave guide, but its half width is independent of its amplitude. In vacuum the signal velocity equals the group velocity. In dispersive media the frequency band width of the signal has to be correspondingly narrow in order to avoid any signal deformation. Physical signals are per se frequency band limited. For example, in fiber optics the band width is about $10^{-4}$ of the infrared carrier frequency. All superluminal experiments discussed in this article are restricted to the signal velocity and neither to the group nor to the phase velocities. The analysis of the experimental data has proven that no signal reshaping took place during tunneling and that all frequency components were equally transmitted. In Ref.~\cite{Steinberg} superluminal tunneling of single photons has been reported. Incidentally, does the measurement of a tunneled particle not represent a signal, which reports with superluminal velocity about the fission and the state of a special nucleus?

The justification for comparing optical experiments with
quantum mechanical tunneling is their mathematical analogy~\cite{Sommerfeld}. The Helmholtz equation is Lorentz invariant whereas the Schr\"odinger equation is Galilei invariant. In the case of evanescent modes with c  $ \rightarrow \infty$ the Helmholtz equation loses its Lorentz invariance.

In classical physics the imaginary part of a wave number k describes a continuous  attenuation of the wave packet with distance. However, in the case of tunneling modes the wave packet after it has entered the barrier is not attenuated during the barrier passage. A wave packet leaves the barrier with the same magnitude as it has entered it, the packet is	elastically scattered at the entrance. A fortiori the incident wave packet knows all about the barrier height and length within the scattering time and is accordingly partially reflected and transmitted.

\begin{figure}[htb]
\includegraphics{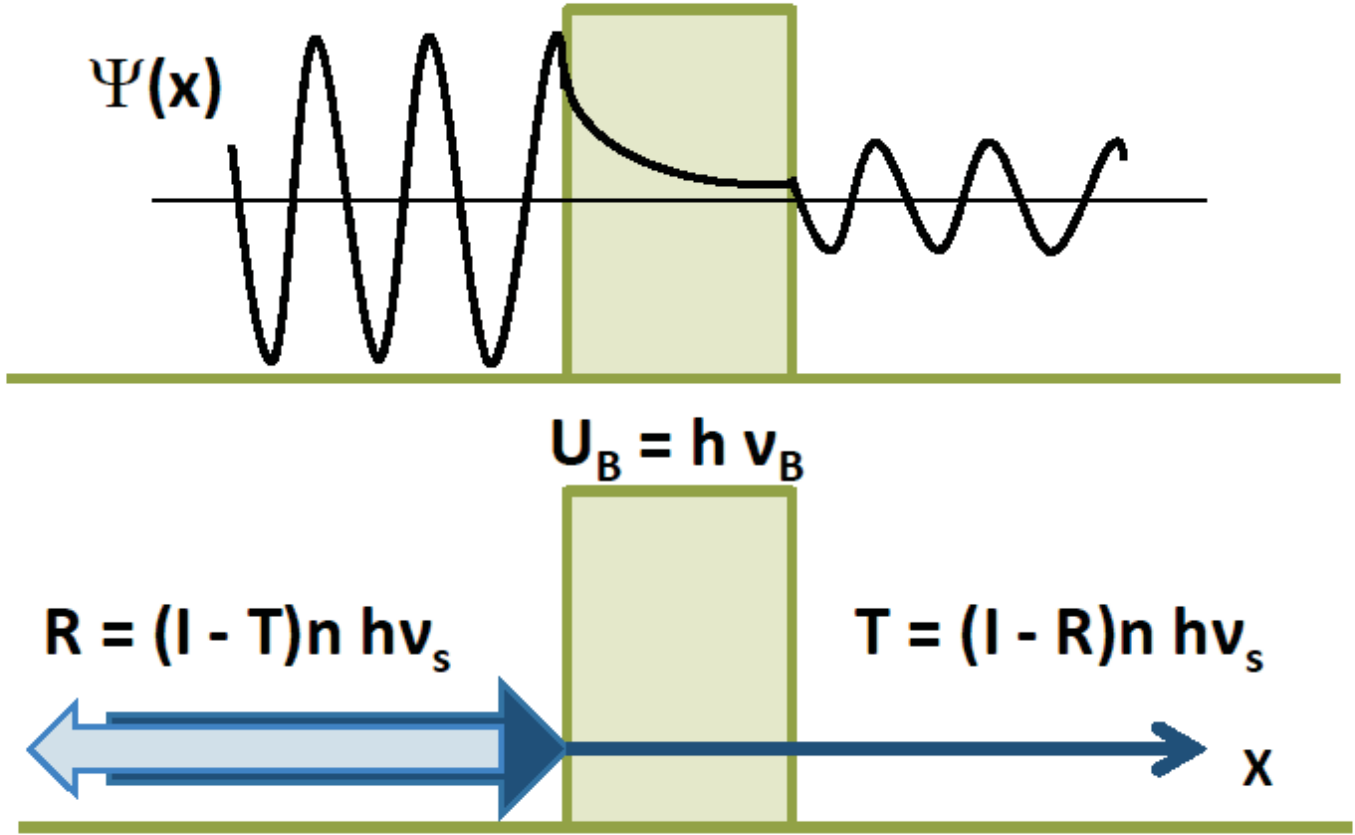}
\caption{Sketch of the wave function $\Psi$ and of the energy of a photon beam I = n h$\nu_s$ incident on barriers with potential energy $U_B$ = h$\nu_B$. Partly reflected energy R = (I - T) n h$\nu_s$ and partly transmitted energy T = I - R. Where $\nu_s$ is the signal (wave packet) frequency, h is the Planck constant, and n represents the number of incident quanta.
 T decreases exponentially with barrier length x  due to reflection R, however,  already at the barrier front.
Inside a barrier T = (I - R)n h$\nu_s$ is constant.
\label{Barriercrossing}}
\end{figure}

This behavior holds for all potential barriers like periodic lattice structures in their forbidden frequency gap (mirrors)~\cite{Steinberg}, in the case of frustrated total reflection at double prisms~\cite{Haibel}, and in undersized wave guides~\cite{Enders}. The tunneling time is always raised at the barrier boundary. The tunneling process is illustrated in Fig.~\ref{Barriercrossing}. This behavior differs from a complex wave number, where due to the imaginary component of the wave number the wave packet continuously attenuates with increasing distance.

Zero time is spent inside a barrier and the wave packet behaves nonlocally, whereas at the barrier front a scattering time takes place of the order of one oscillation of the wave packet's frequency ~\cite{NimtzM,Esposito}. Reflection time and transmission times are identical as a consequence of the nonlocal property of the modes. The boundary
interaction time is independent of the barrier length. This so-called Hartman effect was calculated by Hartman and was first verified with microwaves in 1993~\cite{Hartman,Enders2,NimtzR}.

\section{Evanescent and Tunneling Modes}

Helmholtz and Schr\"odinger equations are mathematical
analogues. From the Helmholtz equation we get the wave number

\begin{eqnarray}
k^2 & = & \frac{n^2 \omega^2}{c^2}\\
k^2 & = &   k_0^2 n^2 \label{A} ,
\end{eqnarray}
where $k_0$ and c are the wave number and the velocity of light in free space respectively. If the refractive index $n$ and thus k
are purely imaginary then the solution is called an
evanescent mode or photonic tunneling.

From the Schr\"odinger equation we get the wave number
\begin{eqnarray}
k^2 & = & \frac{2 m W}{\hbar}(1 - \frac{U}{W})\\
k^2 & = & k_0^2(1 - \frac{U}{W}) \, ,
\end{eqnarray}
where $k_0$ is the wave vector for a
free particle and $\hbar$ the Planck constant.
Particles in regions for which $W < U$, that is, inside the
barrier, are quantum analogs of the evanescent modes~\cite{Sommerfeld}.

\section{Phase Time Approach}
The Wigner phase time delay neglecting dispersion effects equals the signal time delay
\begin{eqnarray}
t_\varphi(\omega) & = & d\varphi(\omega)/d\omega,
\end{eqnarray}
where $\omega$ is the angular frequency and $\varphi$ = k$\cdot$x the phase shift along the distance x. In the case of a purely imaginary wave number k there is no real time expected and indeed not measured~\cite{Pap}. Actually, the phase time approach is used to measure the delay time of electronic and dielectric devices with frequency analyzers nowadays.

\section{The Einstein Energy Relation}

According to the special theory of relativity the
total energy $W$ of a particle is given by the relation
\begin{eqnarray}
W^2 & = & (\hbar k c)^2 + (m_0 c^2)^2 \, \label{energy},
\end{eqnarray}
where  $m_0$ is the rest mass.
Evanescent and tunneling modes with purely imaginary $k$ violate this relation.

Quantum mechanics and classical electrodynamics result in a negative
energy for particles inside a barrier. For evanescent modes this is the electric energy
\begin{eqnarray}
W_{el} & = & \frac{1}{2} \epsilon E^2 V  < 0 \label{energy1}, \label{C}
\end{eqnarray}
where E is the electric field and V is the volume. $\epsilon = n^2$ is the negative dielectric function in the barrier.
Tunneling and evanescent modes violate the Einstein relationship Eq.\ref{energy}.

\section{Nonlocality and Absolute Time}

Tunneling and evanescent modes are nonlocal. They are instantly
spread out with the same amplitude over the whole barrier space. They represent an action at a distance. This follows from experimental results ~\cite{NimtzR,Haibel2} and from theoretical studies~\cite{Ali,Stahlhofen,Carni,Low}.

For $c_{signal}$ $\rightarrow \infty$ we obtain the absolute time of a Galilei-Newton
world in tunneling barriers, whereas finite signal velocities are representative of a local interaction~\cite{Recami}.

\section{Evanescent Modes and Tunneling Particles are not Observables}
Remarkably, evanescent modes like tunneling particles~\cite{NimtzM,Nimtza,Fillard,Merz,Gasio} are not
observable. For
instance, evanescent modes don't interact with an antenna as long
as the barrier system is not perturbed and the evanescent mode is
transformed back into a detectable mode with a real wave number.
An evanescent field does not interact with real fields due to the
imaginary wave number resulting in a refractive index mismatch with the reflection  R = 1.

In the case of quantum mechanical tunneling the particle can be measured only if it is given an energy sufficient to raise it into the classical allowed region~\cite{Merz,Gasio}

\section{Evanescent and Tunneling Modes are Virtual Particles}

It was
pointed out first by Ali \cite{Ali} that virtual photons are those
modes which do not satisfy the Einstein relation Eq.~\ref{energy}.
In the early 1970's several QED calculations showed that
the above mentioned strange behavior of evanescent modes are
adequately described by virtual photons~\cite{Stahlhofen}. The
quantization of evanescent modes by Carniglia and Mandel has shown
that the locality condition is not
fulfilled~\cite{Stahlhofen,Carni}. They concluded that the
commutator of the field operator does not vanish for space-like
separated points.

The experiments have shown that photonic tunneling proceeds with an infinite signal velocity inside the barrier.
A tunneling barrier represents a space of action at a distance.

The same tunneling behavior was observed for phonons and for electrons as demonstrated in Table 1 Ref.~\cite{NimtzM}.

\section{Summary}

As discussed in former articles, the \emph{primitive causality}, i.e. effect follows cause, is
guaranteed~\cite{Nimtz,Nimtza}. Tunneled signals have been
detected at a time shorter than that of vacuum traveled signals. Measured signal velocities were up to 5$\cdot$c~\cite{Longhi,Longhi2,Nimtz5}.
However, due to a signal's product of finite frequency band width
times finite time duration and considering the transmission
dispersion of any barrier, the tunneled signals will begin in the
past but end up in the future~\cite{Nimtz,Nimtza}. That is,
primitive causality applies
even in the case of superluminal signal velocities. The latter is
contradictory to most text books on special relativity, see
e.g.~Refs.~\cite{Fayngold}.  It is usually assumed that a signal has a point like time duration $\Delta t$ $\rightarrow$ 0. This assumption has no physical reality because a wave packet and a signal, even if it informs us about an event in a distant galaxy, follows the relation~\cite{Nimtza,Shannon}
\begin{eqnarray}
\Delta \nu \Delta t & \geq & 1,
\end{eqnarray}
where $\Delta \nu$ and $\Delta t$ are the frequency band width and the time duration of a wave packet. This relation would correspond for $\Delta t$ $\rightarrow$  0 to  an infinite frequency band width and thus according to quantum mechanics to an infinite signal energy. The various mentioned properties run counter of the intuition of special relativity.

In a review article on the experimental proof of quantum teleportation the authors made the statement: \emph{Einstein among many other distinguished physicists, could simply not accept this spooky action at a distance. But this property of entangled states have now been demonstrated by numerous experiments }~\cite{Bouw}. We think this statement applies as well for the strange properties of zero-time barrier space and superluminal signal velocity in tunneling.


\end{document}